\documentclass[aps,prb,twocolumn,reprint]{revtex4-1}
\usepackage{graphicx}
\usepackage{bm}
\usepackage{amssymb}

\begin{document}

\title{Top gating of epitaxial (Bi$_{1-x}$Sb$_x$)$_2$Te$_3$ topological insulator thin films}

\author{Fan Yang}
\affiliation{Institute of Scientific and Industrial Research, Osaka University, Ibaraki, Osaka 567-0047, Japan}
\author{A. A. Taskin}
\affiliation{Institute of Scientific and Industrial Research, Osaka University, Ibaraki, Osaka 567-0047, Japan}
\author{Satoshi Sasaki}
\affiliation{Institute of Scientific and Industrial Research, Osaka University, Ibaraki, Osaka 567-0047, Japan}
\author{Kouji Segawa}
\affiliation{Institute of Scientific and Industrial Research, Osaka University, Ibaraki, Osaka 567-0047, Japan}
\author{Yasuhide Ohno}
\affiliation{Institute of Scientific and Industrial Research, Osaka University, Ibaraki, Osaka 567-0047, Japan}
\author{Kazuhiko Matsumoto}
\affiliation{Institute of Scientific and Industrial Research, Osaka University, Ibaraki, Osaka 567-0047, Japan}
\author{Yoichi~Ando}\email[Electronic mail: ]{y\_ando@sanken.osaka-u.ac.jp}
\affiliation{Institute of Scientific and Industrial Research, Osaka University, Ibaraki, Osaka 567-0047, Japan}

\date{\today}

\begin{abstract}

The tunability of the chemical potential for a wide range encompassing
the Dirac point is important for many future devices based on
topological insulators. Here we report a method to fabricate highly
efficient top gates on epitaxially grown (Bi$_{1-x}$Sb$_x$)$_2$Te$_3$
topological insulator thin films without degrading the film quality. By
combining an \textit{in situ} deposited Al$_2$O$_3$ capping layer and a
SiN$_x$ dielectric layer deposited at low temperature, we were able to
protect the films from degradation during the fabrication processes. We
demonstrate that by using this top gate, the carriers in the top surface
can be efficiently tuned from $n$- to $p$-type. We also show that
magnetotransport properties give evidence for decoupled transport
through top and bottom surfaces for the entire range of gate voltage,
which is only possible in truly bulk-insulating samples.

\end{abstract}

\maketitle

Topological insulators (TIs) are a class of insulators possessing an
energy gap in the bulk and helically spin-polarized gapless states on
the surface.\cite{TI1,TI2,TI3} In the last few years, great efforts and
progress have been made to synthesize high-quality, bulk-insulating TI
materials in various ways.\cite{DopingSingleCrystal1,
DopingSingleCrystal2, DopingSingleCrystal3, DopingSingleCrystal4,
DopingSingleCrystal5, DopingSingleCrystal7, DopingSingleCrystal9,
DopingMBE1, Checkelsky, Kong}. As the materials issues being
solved, the focus and interest of experimentalists start to shift to the
realization of the theoretically proposed TI devices for
applications ranging from spintronics to quantum computing.
\cite{TI1, TI2, TI3} For such devices, an efficient gate
control to tune the Fermi level in a wide range is usually essential,
and already many gating experiments have been done on TI materials by
using different gating techniques. \cite{Checkelsky, Kong, gating1,
gating2, gating3, gating4, gating5, gating6, gating7} Among the various
gating techniques, top gating has an advantage to provide local control
of the chemical potential, which is favorable in many applications.
However, a damage-free method to fabricate efficient top gates on TIs
has not yet been fully developed. 

To achieve efficient top-gate controls of TI materials, there are two main
difficulties. One is to obtain bulk-insulating samples, because the
unwanted bulk carriers bring the Fermi level far away from the Dirac
point and screen the electric field. This difficulty is being solved by
means of chemical doping, in both single crystals
\cite{DopingSingleCrystal1, DopingSingleCrystal2, DopingSingleCrystal3,
DopingSingleCrystal4, DopingSingleCrystal5, DopingSingleCrystal7,
DopingSingleCrystal9, Checkelsky, Kong} and thin films.\cite{DopingMBE1}
The other difficulty is the degradations of the sample surface during
the fabrication processes, especially during the deposition of the
dielectric material. It has been shown that the usual methods to deposit
dielectric materials such as atomic layer deposition (ALD) can cause
heavy damage to the TI surface;\cite{gating2} such a damage would lower
the surface mobility and also introduce a large number of impurity
states to pin the surface chemical potential, making its gate tuning to
be difficult.\cite{gating2} It is thus important to develop a method to
fabricate top gates with minimal damage to the TI surface.

In this Letter, we report our top-gating results on
(Bi$_{1-x}$Sb$_x$)$_2$Te$_3$ topological-insulator thin films grown by
molecular beam epitaxy (MBE). Following Ref. \onlinecite{DopingMBE1},
the composition of the films was optimized for achieving the
bulk-insulating state. We show that a combination of an \textit{in situ}
deposited Al$_2$O$_3$ capping layer and a 200-nm-thick SiN$_x$
dielectric layer deposited by the hot-wire chemical vapor deposition
(CVD) technique at $< 80^{\circ}$C makes it possible to fabricate nearly
damage-free top gate, with which an efficient ambipolar gating is
possible. The mobility and carrier density of both top and bottom
surfaces were estimated by fitting the magnetic-field dependences of the
Hall resistivity, $R_{yx}(B)$, to the two-band model for a range of gate
voltage $V_g$. The mobility of the top surface, 700 -- 1300
cm$^2$V$^{-1}$s$^{-1}$ depending on $V_g$, is found to be always higher
than that of the bottom surface. The weak antilocalization (WAL) effect
was observed in the entire range of $V_g$ and its analysis allows us to
conclude that the top and bottom surfaces are always decoupled in our
films irrespective of $V_g$. Our top gating technique is useful for
realizing TI devices that require both the local control of the
chemical potential encompassing the Dirac point {\it and} a high
mobility of the surface state.

The (Bi$_{1-x}$Sb$_x$)$_2$Te$_3$ thin films were grown on sapphire
(0001) substrates. In this work, the thickness was fixed at 20 nm, for
which the optimized flux ratio between Bi and Sb to obtain
bulk-insulating films was 0.15. After the growth, the sample was
transferred to an attached electron-beam deposition chamber to deposit
4-nm-thick Al$_2$O$_3$ capping layer, during which the sample holder was
cooled by running water to keep the sample temperature below
100$^{\circ}$C. The inset of Fig. 1(b) is an atomic force microscope
(AFM) image of a typical film used in this experiment, a 20-nm-thick
(Bi$_{1-x}$Sb$_x$)$_2$Te$_3$ film capped with a 4-nm-thick Al$_2$O$_3$
capping layer. One can see atomically flat triangles with sharp edges,
indicating that the Al$_2$O$_3$ layer is uniformly deposited. The step
height at the edge of each triangle is about 1 nm, corresponding to the
thickness of the growth unit (quintuple layer) of
(Bi$_{1-x}$Sb$_x$)$_2$Te$_3$.

The films were fabricated into top-gated Hall-bar devices by five steps
described in the following. All metal electrodes of the devices were
made of 50-nm-thick Pd films deposited by magnetron sputtering.

The first step was to pattern the film into a Hall-bar shape with
photolithography and wet etching. Following the wet-etching recipe for
Bi$_2$Te$_3$ films, \cite{WetEtching} we etched our
(Bi$_{1-x}$Sb$_x$)$_2$Te$_3$ films by using the soludion of 1 HCl : 0.8
H$_2$O$_2$ : 8 CH$_3$COOH : 16 H$_2$O (volume ratio); the mass
concentrations of HCl, H$_2$O$_2$, and CH$_3$COOH used for making this
etchant were 36\%, 33\% and 99.7\%, respectively. In the second step,
the resist pattern for metal contacts was defined with photolithography,
and the sample was dipped into 2.4\% tetramethylammonium hydroxide
solution for several minutes to remove the Al$_2$O$_3$ capping layer in
the contact area, followed by Pd deposition and lift-off. The third step
was to deposit the dielectric layer for the top gates. A 200-nm-thick
SiN$_x$ layer was deposited by hot-wire CVD, during which the sample
temperature was kept lower than 80$^{\circ}$C; \cite{Maehashi} since
there is no resist mask for this deposition, the SiN$_x$ layer covers
everywhere, including the Pd contacts on the arms of the Hall bar. The
fourth step was thus to open a window in the SiN$_x$ layer on the top of
each Pd contact; this was done by photolithography and subsequent dry
etching. In the final step, the top-gate electrodes were fabricated by
photolithography, Pd deposition, and lift-off.

\begin{figure}
\includegraphics[width=7cm]{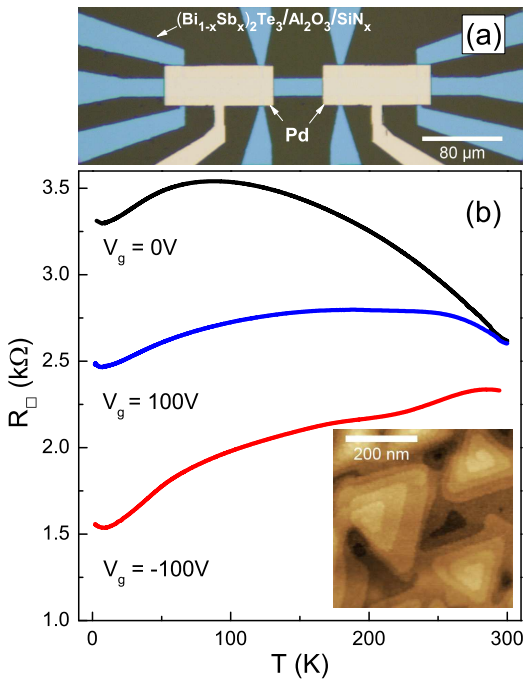}
\caption{(Color online)
(a) Optical microscope image of the device.
(b) Temperature dependences of $R_{\square}$, taken
at $V_g$ = 100 V, 0 V and $-100$ V. Inset: AFM image of a 20-nm-thick
(Bi$_{1-x}$Sb$_x$)$_2$Te$_3$ film capped with 4-nm-thick Al$_2$O$_3$.
}
\label{fig:Fig1}
\end{figure}

The measurements were performed in a $^4$He cryostat with a base
temperature of 1.8 K, by using a standard lock-in technique. The
excitation current was 1 $\mu$A. Using the Hall-bar-shaped samples as
shown in Fig. 1(a), both the longitudinal sheet resistance $R_{\square}$
and the Hall resistance $R_{yx}$ were obtained by four terminal
measurements, as a function of temperature $T$ and magnetic field $B$.
The gate voltage $V_g$ was applied in the range of $\pm100$ V. When the
gate voltage is cycled, we observed a hysteresis in both $R_{\square}$
and $R_{yx}$; this is probably due to impurity states that trap charges.
Hence, to keep consistency in the data, all curves shown here were taken
upon decreasing $V_g$ from +100 V to $-100$ V.

Figure 1(b) shows the $R_{\square}$ vs. $T$ curves measured at three
gate voltages. One can see that the effect of top gating on the
$R_{\square}(T)$ behavior is large; in particular, $R_{\square}$ changes
by more than a factor of two at low temperature. One may notice
that there is a weak 
upturn in all three curves at the lowest temperatures; such an upturn
has been reported for TI thin films in the past, \cite{Liu_eeInt, 
Wang_eeInt, Takagaki_eeInt} and is discussed to be due to 
electron interaction effects. \cite{Pal_eeInt_theory,
Konig_eeInt_theory}

\begin{figure*}
\includegraphics[width=16cm]{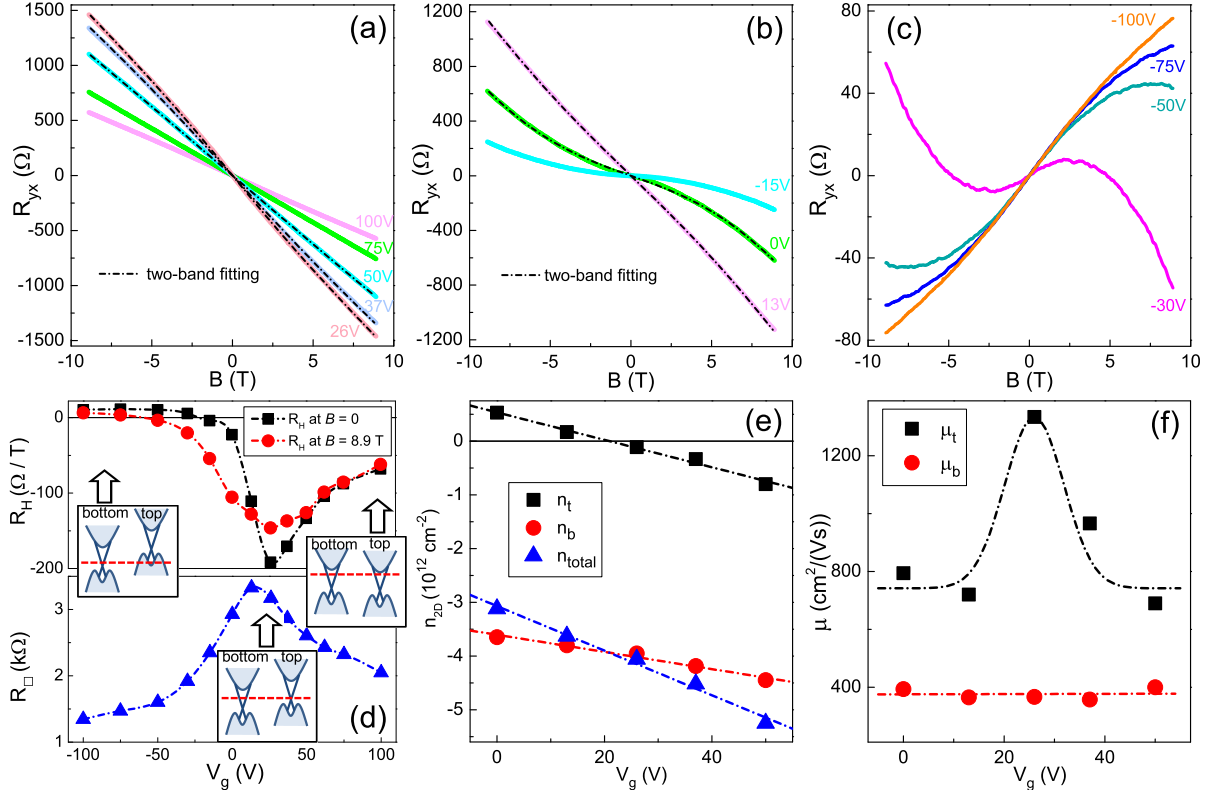}
\caption{(Color online)
(a)-(c) $R_{yx}$ vs. $B$ data at different $V_g$ values. The
dashed-dotted lines in (a) and (b) show fittings to the
two-band model. (d) $V_g$ dependence of $R_H$
(upper panel) and $R_{\square}$ (lower
panel). The insets are schematic band diagrams for three characteristic
regimes. (e) Sheet carrier density vs. $V_g$ for
the top ($n_t$) and bottom ($n_b$) channels, as well as their sum
($n_{\rm total}$). (f) Mobility vs. $V_g$ for top and bottom surfaces.
The data in (e) and (f) are the results of the two-band analysis.
}
\label{fig:Fig2}
\end{figure*}

The $R_{yx}(B)$ curves showed significant and complicated changes when
$V_g$ is varied. In Fig. 2, to present its changes in an comprehensible
manner, we divide the $R_{yx}(B)$ curves measured at various $V_g$ values
into three groups [Figs. 2(a) -- 2(c)] and discuss them separately:

In the first group [Fig. 2(a)], the slopes of the curves are negative,
corresponding to $n$-type carriers. The negative slope increases with
decreasing $V_g$, becoming the steepest at $V_g$ = 26 V. Hence, in this
regime the number of mobile electrons decreases as $V_g$ is reduced.
Below $V_g$ = 50 V, the $R_{yx}(B)$ curves start to show nonlinearity.

In the second group [Fig. 2(b)], while the dominant carriers are still
$n$-type, the slope now decreases with decreasing $V_g$, and the
$R_{yx}(B)$ curves become more nonlinear at lower $V_g$. This suggests
that the chemical potential of the top surface is lowered sufficiently
to cross the Dirac point and $p$-type carriers start to show up on the
top surface; on the other hand, the bottom surface, which feels screened
electric field and is less sensitive to $V_g$, still harbors $n$-type
carriers.

In the third group [Fig. 2(c)], as $V_g$ is lowered into large negative
values, the slope of $R_{yx}(B)$ changes from negative to positive,
indicating a switchover in the dominant carrier type. In this regime,
the density of $p$-type carriers in the top surface steadily increases
with decreasing $V_g$, causing the whole $R_{yx}(B)$ curve to show
positive slope at $V_g$ = $-100$ V.

The $V_g$ dependences of the Hall coefficient $R_H$ and
$R_{\square}$ are summarized in Fig. 2(d). Since the $R_{yx}(B)$ curves
were nonlinear in the majority of the $V_g$ range, we define the slopes
of $R_{yx}(B)$ at $B$ = 0 and 8.9 T as the $R_H$ values at those fields.
The insets in Fig. 2(d) depict the location of the chemical potential at
the top and bottom surfaces for different $V_g$ regimes. At high
positive and negative gate voltages, the carrier type of the top surface
is $n$ and $p$, respectively. However, at $V_g$ = 26 V, $|R_H|$ reaches
a maximum, meaning that the number of mobile carriers becomes minimal at
this $V_g$; such a behavior is expected when the Fermi level of the top
surface hits the Dirac point. The peak in $R_{\square}$ is located at
$V_g$ = 13 V, which is a little different from the maximum position of
$|R_H|$. Such a behavior is frequently reported in gating results of TIs
\cite{gating3, gating4, gating7} and is likely to be due to a difference
in the mobility of the carriers in the top and bottom surfaces.

The observed ambipolar gating suggests that the chemical potential in
our sample is located within the bulk band gap in the majority of the
$V_g$ range, where the main contribution to the transport comes from the
two surfaces channels. Therefore, one would expect that the two-band
model\cite{TI3} to consider two independent surface channels would
describe the $R_{yx}(B)$ behavior. Indeed, as we show below, for the
$V_g$ range of 0 -- 50 V, the fitting of the $R_{yx}(B)$ data to the
two-band model gives reasonable and consistent parameters for the two
surfaces, which reconfirms that it is not necessary to consider an
additional bulk channel. It should be noted, however, that at large
positive $V_g$ values, the $R_{yx}(B)$ curves becomes essentially linear
and featureless, which prohibits us from extracting reliable information
from the two-band analysis. Also, since the top of the bulk valence band
is located close to the Dirac point, \cite{DopingMBE1} at negative
$V_g$ values the Fermi level at the top surface is brought into the bulk
valence band, causing the top surface to harbor conventional inversion
layer (topologically trivial two-dimensional states with $p$-type
carriers) alongside the $p$-type Dirac fermions; this causes the
transport to occur through three channels, and one should not apply
the two-band analysis for $V_g \lesssim$ 0. 

In the two-band model, \cite{TI3} the composite Hall
resistivity is given as 
\begin{equation}
R_{yx}\left(B\right) =
\left(\frac{B}{e}\right)\frac{\left(n_t\mu_t^2+n_b\mu_b^2\right)+B^2\mu_
t^2\mu_b^2\left(n_t+n_b\right)}{\left(n_t\mu_t+n_b\mu_b\right)^2+B^2\mu_
t^2\mu_b^2\left(n_t+n_b\right)^2},
\end{equation}
where $n_t$, $n_b$, $\mu_t$,
$\mu_b$ correspond to the sheet carrier densities and mobilities of the
top and bottom surfaces. Although there are four fitting parameters, one
can put a constraint that the observed zero-field sheet resistance 
$R_\square|_{B=0}$ must
be consistent with $1/(e n_t\mu_t+e n_b\mu_b)$, which reduces the number
of free parameters to three. \cite{TI3}

The fittings of the data for $V_g$ = 0 -- 50 V to Eq. (1) are shown in
Figs. 2(a) and 2(b), and the obtained parameters are shown in Figs. 2(e)
and 2(f). One can see that both $n_t$ and $n_b$ present linear $V_g$
dependences, and the slope is larger for $n_t(V_g)$ (i.e., the top
surface is more efficiently gated), which is reasonable and reassures 
our assumption of two transport channels. The sign change
in $n_t$ at $V_g \approx$ 26 V indicates that the Fermi level of the top
surface hits the Dirac point at this $V_g$.

In Fig. 2(f), one can see that $\mu_t$ presents a peak value of 1300
cm$^2$V$^{-1}$s$^{-1}$ near the Dirac point, and elsewhere $\mu_t$
remains around 750 cm$^2$V$^{-1}$s$^{-1}$. Such an enhancement of the
mobility near the Dirac point is often observed in graphene and is due
to the vanishing of the phase space available for scattering.
\cite{graphene1,graphene2} One may also notice that the bottom-surface
mobility $\mu_b \simeq$ 375 cm$^2$V$^{-1}$s$^{-1}$ is much lower than
$\mu_t$, meaning that the electron scattering is stronger at the
interface with sapphire substrate than at the interface with Al$_2$O$_3$
capping layer. We have also measured the $R_{yx}(B)$ curves in
capping-layer free (Bi$_{1-x}$Sb$_x$)$_2$Te$_3$ films without any
patterning nor gate fabrication, and in such pristine films, $\mu_t$
was found to be typically $\sim$800 cm$^2$V$^{-1}$s$^{-1}$. This
indicates that the Al$_2$O$_3$ capping layer is indeed effective in
keeping the top-surface mobility from degradation. In passing, we
observed that the capping layer tends to dope $p$-type carriers to the
surface, causing the top surface band to be slightly bent up.

\begin{figure}
\includegraphics[width=8.5cm]{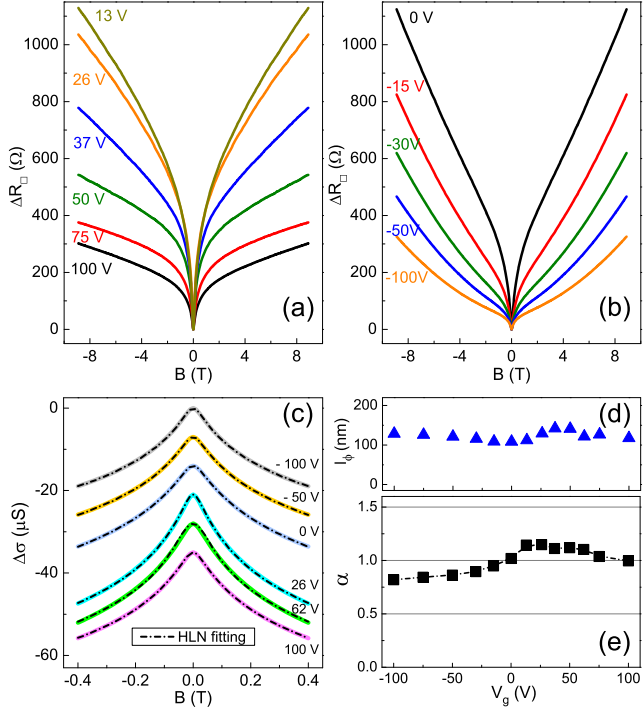}
\caption{(Color online)
(a)-(b) Magnetoresistance at different gate voltages; (a) is for $V_g
\ge$ 13 V and (b) is for $V_g \le$ 0 V. (c) Magnetoconductance at
representative $V_g$ values; the curves are shifted vertically for
clarity. The dashed-dotted lines are the fittings to Eq. (2). (d)-(e)
$V_g$ dependences of (d) $l_\phi$ and (e) $\alpha$ obtained from the
fittings. 
}
\label{fig:Fig3}
\end{figure}

Now we discuss the WAL effect in our films to see if the top and bottom
surfaces are decoupled. In Figs. 3(a) and 3(b), we plot the
magnetoresistance $\Delta R_{\square}(B)$ [$\equiv R_{\square}(B) -
R_{\square}|_{B=0}$] for various $V_g$ values, and the curves are
divided into two groups. In the first group [Fig. 3(a)], the
magnetoresistance presents negative curvature throughout the
magnetic-field range, while in the second group [Fig. 3(b)] the
magnetoresistance seems to contain a parabolic background. Since the
boundary between the two groups is $V_g$ = 13 V where $R_{\square}$
reaches the maximum, the change in the background in the
magnetoresistance appears to be correlated with the change in the
carrier type. However, the origin of this intriguing behavior is not
known at the moment.

Despite the change in the high-magnetic-field background, the cusp
structure in the low-magnetic-field part of $R_{\square}(B)$ 
can always be reliably analyzed as the WAL behavior. The WAL in TIs has
been intensively studied both theoretically \cite{WAL1, WAL2} and
experimentally \cite{gating3, gating4, gating8, gating9, WAL3, WAL4,
WAL5}. The peculiar magnetic-field dependence of the electrical
conductivity $\sigma$ associated with WAL is described by the
simplified Hikami-Larkin-Nagaoka (HLN) formula 
\begin{equation}
\Delta\sigma\left(B\right)=-\alpha\frac{e^2}{\pi
h}\left[\Psi\left(\frac{1}{2}+\frac{\hbar}{4e l_{\phi}^2 B}\right)-
\ln \left(\frac{\hbar}{4e l_{\phi}^2 B}\right)\right],
\end{equation}
where $\Psi(x)$ is the digamma function and $l_\phi$ is the phase
coherence length. The parameter $\alpha$ is presumably 0.5 when there is
only one transport channel, and it is doubled when there are two
channels. We have calculated $\sigma(B)$ from the data of
$R_{\square}(B)$ and $R_{yx}(B)$ to do the proper matrix inversion.

As shown in Fig. 3(c) for representative $V_g$ values, all the $\Delta
\sigma(B)$ curves are well fitted with the HLN formula. The parameters
extracted from the fittings are plotted in Figs. 3(d) and 3(e). The
$l_\phi$ value [Fig. 3(d)] is essentially independent of $V_g$, and its
average value of $\sim$120 nm is much larger than the film thickness.
More importantly, the $\alpha$ value [Fig. 3(e)] is found to be close to
1 throughout our $V_g$ range, \cite{note} pointing to the existence of two
independent channels that are presumably the top and bottom surfaces. It
has been elucidated that in TI thin films, residual bulk carriers couple
the top and bottom surfaces through the bulk to produce a single
diffusive transport channel for the WAL effect, which results in the
$\alpha$ value of 0.5. \cite{gating1, gating3, gating4, WAL4} The
$\alpha$ value close to 1 has been observed only in two situations: (i)
a depletion layer is formed in the film due to a strong band bending at
high gate voltages to electrically isolate the gated surface,
\cite{gating3, gating4} or (ii) films are relatively thick (which avoids
the hybridization of the top and bottom surface states) and bulk
carriers are negligible. \cite{WAL4} In this regard, the $\alpha$ value
in our film is nearly independent of $V_g$, which rules out the former
possibility and points to the latter. Thus, the analysis of the WAL
behavior corroborates the conclusion that our films are truly
bulk-insulating.

In summary, we developed a damage-free method to fabricate efficient top
gates on (Bi$_{1-x}$Sb$_x$)$_2$Te$_3$ topological-insulator thin films.
The carrier type in the top surface can be tuned from $n$ to
$p$, and their mobility reaches 1300 cm$^2$V$^{-1}$s$^{-1}$
near the Dirac point. The magnetotransport properties give evidence that
our films are bulk insulating with decoupled top and bottom surfaces.
The top gating method developed here would be useful for the
realization of future devices based on TIs.


We acknowledge M. Kishi for technical assistance and Nanotechnology Open
Facilities in Osaka University for nano-fabrication facilities. This
work was supported by JSPS (KAKENHI 24740237, 24540320, 25400328, and
25220708), MEXT (Innovative Area ``Topological Quantum Phenomena"
KAKENHI), and AFOSR (AOARD 124038).



\begin{thebibliography}{10}
\bibitem{note} 
Following Ref. \onlinecite{Wang_eeInt}, one can estimate
the possible impact of the electron interaction effects on the WAL
analysis; in our case, the screening parameter $\tilde{F}_{\sigma}$
estimated from the $\sigma(T)$ behavior is only $\sim$0.1, introducing 
an error in $\alpha$ of only up to 10\%.
\end{thebibliography}
\end{document}